\documentclass[12pt]{article}

\usepackage{scicite}
\usepackage{graphicx}
\usepackage{times}
\usepackage{pdfpages}

\newenvironment{sciabstract}{%
\begin{quote} \bf}
{\end{quote}}

\title{Strongly interacting dipolar-polaritons}

\author
{Itamar Rosenberg,$^{1}$ Dror Liran,$^{1}$  Yotam Mazuz-Harpaz,$^{1}$\\ Kenneth West,$^{3}$ Loren Pfeiffer,$^{3}$ Ronen Rapaport$^{1,2\ast}$ \\
\\
\normalsize{$^{1}$Racah Institute of Physics, The Hebrew University of Jerusalem, Jerusalem 91904, Israel}
\\\normalsize{$^{2}$The Applied Physics Department, 
The Hebrew University of Jerusalem, Jerusalem 91904, Israel}\\
\normalsize{$^{3}$Department of Electrical Engineering, Princeton University, Princeton, New Jersey 08544, USA}\\
\\
\normalsize{$^\ast$Corresponding author; E-mail:  ronenr@phys.huji.ac.il }
}

\date{}

\begin{document} 


\baselineskip24pt


\maketitle


\begin{sciabstract}
Exciton-polaritons are mutually interacting quantum hybridizations of confined photons and electronic excitations. Here we demonstrate a system of optically guided, electrically polarized exciton-polaritons ('dipolaritons') that displays up to 200-fold enhancement of the polariton-polariton interaction strength compared to unpolarized polaritons. The magnitude of the dipolar interaction enhancement can be turned on and off and be easily tuned over a very wide range by varying the applied polarizing electric-field. The very large interaction strengths and the very long propagation distances of these fully-guided dipolaritons open up new opportunities for realizing complex quantum circuitry and quantum simulators, as well as topological states based on exciton-polaritons, for which the interactions between polaritons need to be large and spatially or temporally controlled. The results also raise fundamental questions on the origin of such large enhancements.
\end{sciabstract}



Exciton-polaritons are quantum superposition states (quasi-particles) of optically confined light and electronically confined excitations (excitons) in semiconductor quantum structures. Polaritons have very low effective masses due to their photonic part and significant particle interactions due to their excitonic part. This combination of properties yielded observations on many collective quantum phenomena\cite{imamoglu_quantum_1996,dang_stimulation_1998,senellart_nonlinear_1999,kasprzak_bose-einstein_2006,balili_bose-einstein_2007,lagoudakis_quantized_2008,roumpos_single_2011,utsunomiya_observation_2008,amo_collective_2009,DengPolaritonLaser} and it makes polaritons appealing for developing non-linear optical components on the quantum level. Indeed it has been recently suggested theoretically that quantum simulators, similar to those demonstrated with cold atoms, can be constructed using quantum condensates of exciton polaritons in lattices \cite{hartmann_quantum_2016,amo_exciton-polaritons_2016}. Furthermore, recent demonstrations of low loss, high-velocity slab-waveguided exciton polaritons \cite{walker_dark_2017,walker_ultra-low-power_2015,rosenberg_electrically_2016,ciers_propagating_2017} have opened up possibilities for demonstrating real functional polariton-based quantum circuitry on an optical chip, in a parallel effort to the one based on Rydberg atoms \cite{saffman_quantum_2010,saffman_quantum_2016,RydbergSinglePhotonTransistor,kubler_coherent_2010,weimer_rydberg_2010}, and to realize interacting topological polaritons ('topolaritons') as was suggested recently \cite{Topolaritons}.

Major efforts are on way towards a realization of proposals for quantum devices using polaritons, however the main obstacle for realizing such quantum systems is that they usually require strong interactions between only two polaritons, which is necessary to achieve entanglement and logic operations for quantum circuitry, and sufficient on-site interactions for quantum lattice simulators. Since the interactions between polaritons stem from the excitonic part of the quasi-particles, and since excitons are neutral, unpolarized particles, these interactions are weak and short ranged, thus one needs to squeeze polaritons into very tiny, deep sub-micron areas to achieve sufficiently strong interactions. Such squeezing of optical waves is a hard challenge, and thus only classical optical simulators and non-linear polariton effects involving many particles has been demonstrated to date  \footnote{We note that a recent work reported evidence for large interactions between unpolarized polaritons \cite{noauthor_direct_nodate}, which is in contrast to previous reports.}. Furthermore, unlike in atomic systems, where interactions can be tuned effectively, in current polariton systems the interaction strength is completely determined by the system parameters and is therefore uncontrollable externally, neither in time nor space. Such fixed interactions may strongly limit applications where external control and tuning of the interaction strength is necessary.  

On the other hand, it is by now well established that while neutral unpolarized excitons interact weakly, excitons that carry an electrical dipole moment  (sometimes known as dipolar excitons), display much larger interactions, manifested by very large energy blue-shifts of their emission line, and large collisional broadening\cite{shilo_particle_2013, KobiShual,stern_exciton_2014}. These interactions stem from the long-range dipole-dipole interactions, $V_{dd}(r)\simeq e^2d^2/(\varepsilon r^3)$, rather than the contact exchange interactions of unpolarized excitons, thus significantly increasing the scattering cross-section, as is evident by recent observations of dipolar spatial correlations of dipolar excitons \cite{shilo_particle_2013,KobiShual,stern_exciton_2014}, dipolar blockade between ensembles \cite{high_control_2008} and between only two excitons \cite{schinner_confinement_2013}. Adopting dipolar interactions into polariton systems by realizing dipolar-polaritons could bridge the "interaction gap" of current polariton systems. However, dipolar excitons are usually realized with the electron and the hole separated into two adjacent quantum wells (a double quantum-well geometry), which diminish their coupling to light and thus inhibits strong coupling and the formation of polaritons. Therefore realizing dipolaritons with a double quantum well structure is very challenging \cite{cristofolini_coupling_2012}. 

Recently we have shown that slab-waveguide dipolaritons can be realized with only a single quantum well geometry, circumventing the strong-coupling challenge and allowing a full electrical control of the electric dipole moment of the polaritons\cite{rosenberg_electrically_2016}. In this paper, we demonstrate a system of truly two-dimensionally guided, low loss, electrically-polarized exciton-dipolaritons with an electrically tunable dipole moment. We show compelling evidence for up to two-orders of magnitude enhancement of the interaction strength between dipolaritons compared to unpolarized polaritons. The dipolar nature of the interaction is demonstrated through the linear dependence of the energy blue-shifts of the polariton signal and the quadratic dependence of the polariton-polariton scattering rate on the dipole size. The magnitude of the dipolar interaction enhancement can be easily controlled over a very wide range by varying the applied polarizing electric-field. The measured interaction strengths are shown to be in the range required for realistic 2-polariton gates. This, together with the very long propagation distances of these dipolaritons, can open up new opportunities for realizing quantum circuitry based on exciton-polaritons.
\begin{figure}[tbp]
  \centering
  \includegraphics[width=0.6\textwidth]{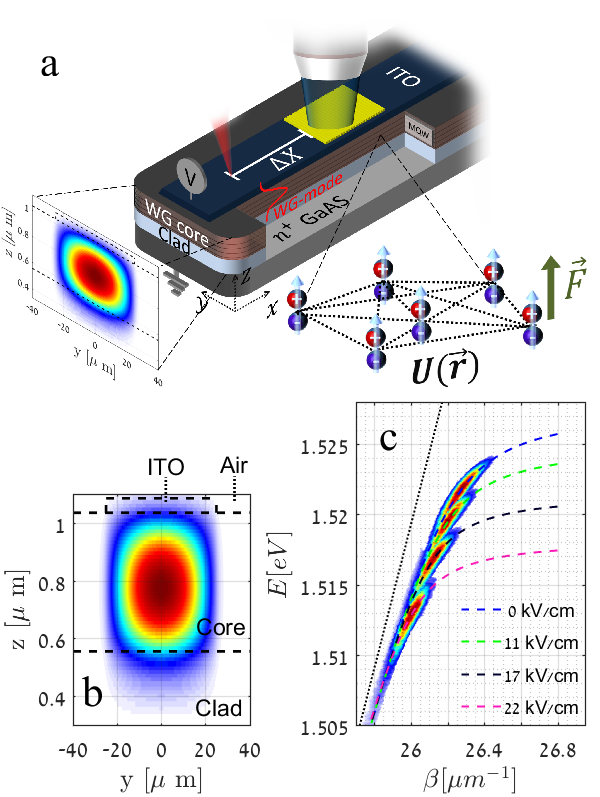}
  \caption{\textbf{Electrically polarized waveguide polaritons with mutual dipole-dipole interactions}. (a) An illustration of the system and experimental method. A non-resonant excitation through a thin ITO channel is used to excite confined polaritons which propagate along the waveguide channel with a propagation constant $\beta$. A gold grating coupler with a periodicity of $240\:nm$ located $\Delta X$ from the excitation spot is used to couple the signal out into the acquisition system. When a voltage is applied to the channel, the polaritons become electrically polarized, acquiring a dipole moment. Polaritons then interact with each other via dipole-dipole interactions. (b) Calculated field distribution of the 1st waveguide mode confined by the ridge waveguide structure. (c) Overlaid PL measured at four different values of $F$, all fitted to the coupled oscillators model. The black dotted line is the bare photon mode. The red-shift of the dispersion is a result of the quantum-confined Stark-shift that polarizes the excitons (full images of all the four corresponding dispersions can be found in the Supplemental material).}
    \label{fig:fig1}
\end{figure}

\begin{figure}[tbp]
  \centering
  \includegraphics[width=0.5\textwidth]{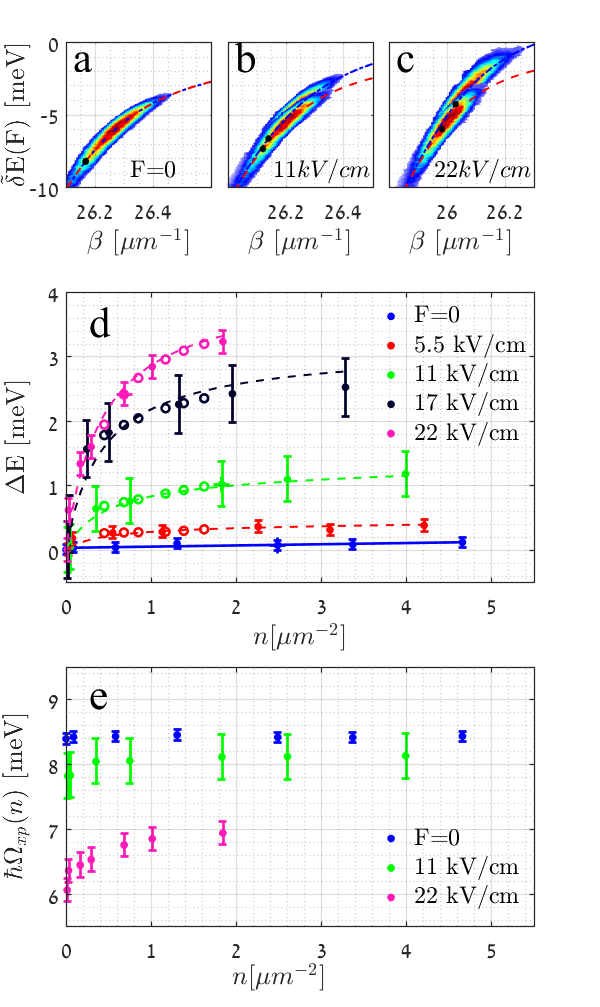}
  \caption{\textbf{Strong dipolar interactions between polaritons}. (a-c) Overlaid polariton PL dispersions measured with low and high excitation powers under different values of $F$. The dashed lines are fits to the coupled oscillators model and the black dots mark polaritons with $\chi_x=1/2$. The energy blue-shift which increases with increasing electric field is an indication for the dipolar interaction between polaritons (full separate images of all the six corresponding dispersions can be found in the Supplemental material). (d) The energy shift of the  polariton mode as a function of the polariton density for different values of $F$. The solid blue line is a linear fit for the case of $F=0$, and the dashed curves are the fits to the theoretical model given in the text. (e) The dependence of $\Omega_{xp}$ on $n$ for various values of $F$. An increase in $n$ results in an energy blue-shift which in turn tends to increase the \textit{optical} dipole moment of the exciton leading to an increase of $\Omega_{xp}$. The effect is stronger as $F$ increases.} 
    \label{fig:fig2}
\end{figure}

The sample structure and the experimental method are plotted in Figure \ref{fig:fig1}(a). The sample is constructed in a waveguide (WG) geometry to support guided polariton modes, made of waveguided photons strongly interacting with excitonic transitions of a set of twelve $20$nm wide quantum wells (QWs) positioned inside the WG core\footnote{The core of the optical waveguide consists of a set of 12 $20$nm wide Al$_{0.4}$Ga$_{0.6}$As/GaAs/Al$_{0.4}$Ga$_{0.6}$As QWs and a $500$nm bottom clad layer made of Al$_{0.8}$Ga$_{0.2}$As.}. The top clad layer is made of a deposited $50$nm thick, $50\mu$m wide transparent and conductive ITO channel. This ITO channel defines a ridge waveguide structure, supporting guided modes which are confined both vertically and laterally, yielding two-dimensional optical confinement for the polaritons, as depicted in Figure  \ref{fig:fig1}(b). The ITO layer also serves as a top electrode, to which voltage can be applied with respect to the $n^+$ doped GaAs back gate, resulting in a constant electric field ($F$) directed perpendicular to the QWs' plane (see Figure \ref{fig:fig1}(a)). The electric field induces a quadratic Stark energy red-shift of the exciton resonance \cite{bastard_variational_1983}, leading to a red-shift of all the polaritonic branches at the anti-crossing point. Furthermore, The applied field induces a polarization of the electron and hole wavefunctions along the z-direction, resulting in a net electric dipole moment of the excitons and thus of the WG-polaritons \cite{rosenberg_electrically_2016}. Both the energy of the WG-polaritons and their dipole moment are controlled by the applied field. 
Our experiments utilize the fact that the propagation velocity of waveguided polaritons is four orders of magnitude larger than the typical propagation velocity of dipolar excitons  (~30 $\mu$m/ps for the polaritons compared to ~1-10$\mu$m/ns for excitons) \cite{snoke_long-range_2002,gartner_drift_2006}. This fact offers a "built-in" mechanism to separate the polaritons from the exciton cloud: at each measurement, our focused laser pulse ($\lambda$=775 nm , duration$\simeq$280 ps) excites WG-polaritons that propagate along the WG from a specific distance $\Delta X$ towards a metallic grating that couples the guided polariton signal out to the measuring optical system (typically $\Delta X$ is from tens to hundreds of microns). The pulse is synchronized with a $10$ ns exposure window of a gated intensified CCD camera (PI-MAX) which images the emission in the Fourier-plane of the grating out-coupler, thus recording the angular resolved dispersion of the signal emitted from the readout area. This type of remote excitation scheme separates the populations of the fast polaritons from that of the slow excitons and thus guarantees that only polaritons are present at the measurement area during the exposure. During all the experiments described here, the sample was positioned inside a cold finger cryostat which maintained a temperature of $\sim 5^\circ K$.   Figure \ref{fig:fig1}(c) shows a set of measured WG-polariton dispersions (showing the polariton energy versus $\beta$, the propagation wave-vector of the polaritons) under various applied electric fields $F$. The effect of the stark shift on the polaritonic dispersion is clearly seen.
Next, we turn to investigate the impact of the induced electric dipole on the polariton-polariton interactions. This can be done by monitoring the changes in the polariton dispersion under variations of the polariton densities (controlled by the excitation power) and variations of the polariton electric dipole (controlled by the applied voltage). Initial evidence for dipolar induced polariton-polariton interactions can be seen in Figure  \ref{fig:fig2}(a-c). Here we placed our excitation spot at distance of $\Delta X=65\mu m$ from the grating out-coupler under different excitation powers and for different values of $F$. Figure \ref{fig:fig2}(a-c) shows overlaid measured polariton dispersions acquired for two excitation powers (low and high) for three different values of $F$ \footnote{full images of  all the corresponding dispersions can be found in the Supplemental material.}. Since the bare exciton energy and thus the polariton energy red-shift with increasing $F$, and since we want to plot all dispersions on the same energy scale, we used an energy scale relative to the bare exciton energy, $\tilde{\delta}E(F)=E-E_x(F)$.
%
While there is only a tiny energy blue-shift of the dispersion when $F=0$  (Figure \ref{fig:fig2}(a)) it can be clearly seen that the polaritonic dispersion blue-shifts in energy with increasing excitation power when $F>0$, and that this energy blue-shift increases with increasing $F$ (Figure \ref{fig:fig2}(b,c)). More quantitative analysis can be acquired by fitting these measured dispersions to the mode dispersions calculated using a full electromagnetic numerical model. The comparison between the measured dispersions and the numerically calculated ones are shown in the supplemental material. Another useful analysis method is a fit to the coupled oscillator model which describes the strong coupling between the exciton and the waveguide mode. These fits are shown by the dashed lines in Figure  \ref{fig:fig2}(a-c)). From those fits, we extract the energy of polaritons for any given excitonic fraction, $\chi_x$. 
Energy variations of the lower polariton mode can result from two mechanisms. The first is due to exciton-exciton interactions that shift the excitonic part of the polariton and thus the whole polariton dispersion. The second is a variation in the optical coupling of the exciton to the photon (due to screening for example) which in turn changes the Rabi splitting $\hbar\Omega_{xp}$, and can also result in an apparent blue-shift of the lower polariton dispersion. It is thus vital to unambiguously identify the source of the observed energy blue shift. Importantly, while such identification might be non-trivial in microcavities,  in WG structures it is quite straightforward: while a blue-shift arising from exciton-exciton interactions is accompanied by a shift in the parallel momentum of the polariton dispersion, a blue-shift that results from mere variations of $\Omega_{xp}$ occurs without such momentum shift. Therefore, by following the variations with $n$ of $\beta(\chi_x=1/2)$  (marked by the black dots in Figure \ref{fig:fig2}(a-c))\footnote{Note that $\beta(\chi_x=1/2)$ is by definition identical to the parallel momentum of the crossing point between the dispersions of the bare modes.}, one can single out the contributions of the interaction-induced blue-shift of the exciton part and the contributions from variations in $\Omega_{xp}$, as is detailed in the supplemental material. In Figure \ref{fig:fig2}(d) we plot the interaction-induced blue-shift energy of the $\chi_x=1/2$ point of the lower polariton branch, $\Delta E=E(n)-E(n=0)$, as a function of the observed polariton density $n$, under the out-coupler, for different values of $F$ and in Figure \ref{fig:fig2}(e) we plot the change in $\Omega_{xp}$ as function of $n$ for different values of $F$  
 \footnote{The polariton density was estimated by carefully counting the photons emitted through the grating and using the proper calibration of losses and cloud size. For more details on the calibration procedure please refer to the supplemental material.}.
%

\begin{figure}[tbp]
  \centering
  \includegraphics[width=0.6\textwidth]{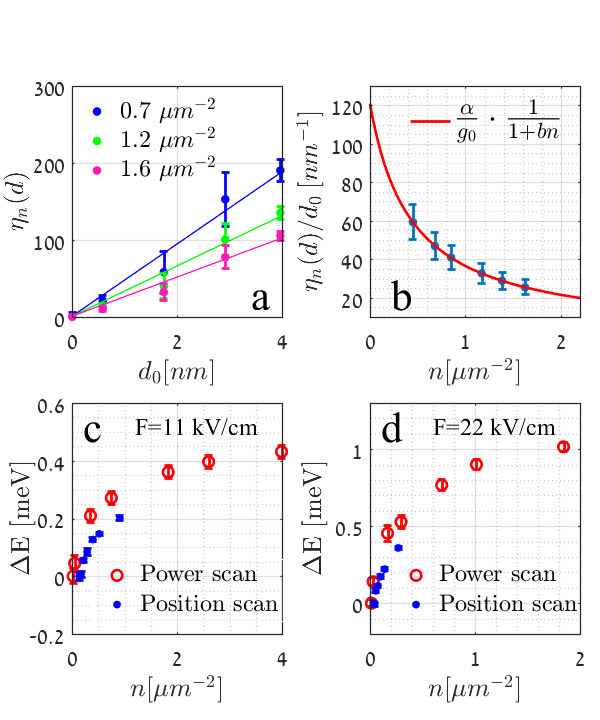}
  \caption{\textbf{The large enhancement of the polariton-polariton interactions}. (a) The interaction enhancement factor $\eta_n(d)$ for various densities (interpolated, marked by the open circles in Figure \ref{fig:fig2}(d)) plotted as a function of $d_0$. Interaction enhancements of up to almost 200-fold are observed for the highest $F$ values. (b) The dependence of $\eta_n(d)\mathbin{/}d_0$ on $n$ fitted to the theoretical model described in the text. (c,d) A comparison between the energy shift measured with a fixed excitation distance $\Delta X$ and an increasing excitation power (red circles) and the energy shift measured with a fixed excitation power and decreasing excitation distance $\Delta X$ for two values of $F$. Here the energy was measured at a constant $\beta$ value (the difference in the magnitudes of $\Delta E (\beta)$ and $\Delta E(\chi_x=1/2)$ is explained in the supplemental material). The similarity between the two experiments suggests that the measured energy blue-shifts are a result of local polariton interactions at the readout point.}
    \label{fig:fig4}
\end{figure}

%
The case of $F=0$ corresponds to unpolarized polaritons while for $F>0$ the polaritons are dipolar (dipolaritons), with increasing dipole moment. This blue-shift energy is the result of polariton-polariton interactions. A small blue-shift is observed for unpolarized polaritons as their density is increased (blue dots in Figure \ref{fig:fig2}(d)). This blue-shift energy is attributed to the short-range, non-dipolar polariton-polariton interactions. From the fitted linear dependence $\Delta E=g_0n$ (solid blue line in Figure \ref{fig:fig2}(d)) the interaction strength $g_0$ can be extracted: $g_0=18\pm 8 \mu eV \mu m^2$. This value is similar to the $g_0=25-37\mu eV \mu m^2$ reported by Walker et al.\cite{walker_dark_2017} for unpolarized WG-polaritons with narrow InGaAs QWs and to the $g_x=30\mu eV \mu m^2$ reported by Rodriguez et al.\cite{rodriguez_interaction-induced_2016} for pure excitons in InGaAs QW.
This agreement reinforces the validity of our polariton density calibration.
A large increase of $\Delta E$ values is observed for the cases where $F>0$, compared to the $F=0$ case. For any tested density, $\Delta E$ is larger for larger $F$'s. This substantial enhancement of $\Delta E$ as $F$ increases is attributed to the contribution of dipole-dipole interactions between dipolaritons, which as can be seen are much stronger than the non-dipolar interactions.
The reduction of $\Omega_{xp}$ with $F$ results from the reduction in the overlap between the wavefunctions of the hole and the electron as $F$ increases\cite{rosenberg_electrically_2016}. As $n$ increases, we find that $\Omega_{xp}$ also increases. This is expected as the blue-shift tends to reverse the field-induced $e$-$h$ separation in the QW and thus $\Omega_{xp}$ increases. We note that we do not observe a significant reduction in $\Omega_{xp}$ due to phase-space filling or carrier screening\cite{huang_carrier_1990}.    In order to verify the dipolar nature of the interactions and quantify their magnitude, we consider the dependence of $\Delta E$ on $F$ for a given constant density. The interaction energy in a dipolar gas is expected to be proportional both to the density of the dipoles and to their dipole length. Thus we can expect the following expression for $\Delta E$:
\begin{equation}\label{Eq:blue-shift}
 \Delta E(d,n)=g_0n+g_d(n)n=g_0n+\alpha d(n)n
\end{equation}
Where $g_d$ is the dipolar interaction strength that characterizes the dipolariton interactions and $d(n)$ is the dipole length which also depends on $n$ \cite{rosenberg_electrically_2016}.  By dividing Eq.\ref{Eq:blue-shift} by $g_0n$ we get an expression for the enhancement factor $\eta(n,d_0)$ which quantifies by how much the interaction strength has increased from the unpolarized case (here $d_0$ is the dipole length for $n=0$):
\begin{equation}\label{Eq:enhancement}
 \eta_n(d_0)=\frac{g_d(d_0,n)}{g_0}=\frac{\alpha d(d_0,n)}{g_0}
\end{equation}
In Figure  \ref{fig:fig4}(a) we plot $\eta$ as function of $d_0$ for three different polariton densities.
%
%
We find that $\eta$ increases linearly with $d_0$, and that huge interaction enhancement factors reaching almost $200$ can be achieved, yielding a maximal measured $g_d\sim 2.4\:meV \mu m^2$. This value is much higher than ever reported for polariton systems as far as we know. 
The linear dependence on the dipole length verifies the dipolar nature of the polariton-polariton enhanced interactions. Moreover, The slopes of the graphs in Figure \ref{fig:fig4}(a) represent the magnitude $\eta/d_0$, from which both $\alpha$ and $d(n)$ can be deduced. In the \textit{supplemental material}, we derive a simple model which shows that the dipole moment is expected to reduce with increasing density as $d(n)=d_0/(1+b n)$, where $b$ is a constant. In Figure \ref{fig:fig4}(b) we plot the slopes as a function of $n$ together with a fit to $\alpha/(g_0(1+bn))$. We find a good agreement between the experiment and the functional form of $d(n)$. We deduce $\alpha\mathbin{/}g_0=122\pm 2\:nm^{-1}$ and $\alpha=2.2\pm 1.0\: eV\mu m$. These extracted values and the above model for $d(F,n)$ are then plugged back into Eq. \ref{Eq:blue-shift}, and are compared to the measured dependence $\Delta E(d_0,n)$, where a best fit is done only within their extracted error limits. The prediction of Eq. \ref{Eq:blue-shift} fits well the experimental results, as is seen by the dashed lines in Figure \ref{fig:fig2}(d). This confirms the self-consistency of our model and analysis.
%

\begin{figure}[tbp]
  \centering
  \includegraphics[width=0.5\textwidth]{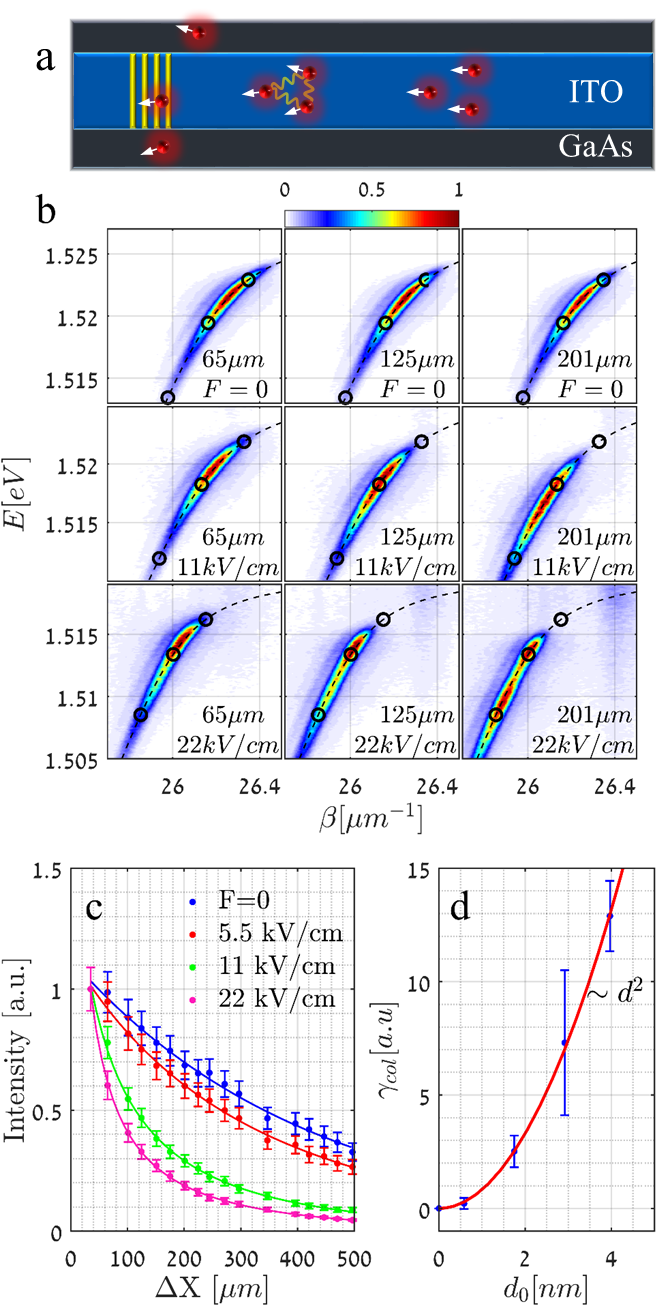}
  \caption{\textbf{Enhanced dipolar-induced polariton scattering} (a) A schematic illustration of a dipolar induced polariton  scattering process leading to a loss of propagating polaritons  in the waveguide. (b) Measured dispersion from various excitation distances and different values of $F$. Each row represents a constant $F$ while each column represents a different excitation distance. The dashed lines are fits to the coupled oscillators model and the circles represent the points in which $\chi_x$ is 1/4, 1/2 and 3/4. A clear redistribution of polariton population is observed for polaritons under an applied electric field. (c) The decay of the emitted intensity with respect to the distance of excitation for different values of $F$. The solid lines are the fits acquired from Eq.\ref{Eq:solution}. (d) $\gamma_{col}$ as a function of the dipole length together with a quadratic fit.}
    \label{fig:fig3}
\end{figure}

Such strong polariton-polariton interactions, observed through the increase of the mean energy of the polariton system, are also expected to have a strong enhancement of polariton-polariton scattering processes. Elastic scattering of polaritons should lead to a redistribution of the polariton population in momentum space \cite{Parametric1,parametric2}. For the case of polaritons in a WG, all having similar propagation constants along the WG direction and small quantized momenta parallel to the propagation direction, these elastic collisions should result in a significant scattering out of the guided modes into leaky, non-guided modes, and thus to additional propagation loss. Inelastic scattering processes should lead to a redistribution of the polariton population along the dispersion. This has motivated  us to look at the propagation dynamics of the dipolaritons. In Figure \ref{fig:fig3}(a-i) we plot the measured polariton emission from the out-coupler for three different excitation distances $\Delta X$, and three different values of $F$. On each plot we show the fitted coupled oscillator model and three points which mark the locations of $\chi_x=3/4$, $\chi_x=1/2$ and $\chi_x=1/4$ respectively.
It is evident that as $F$ increases, there is an increasing "motion" of the apparent polariton population towards lower energies for growing $\Delta X$. This indicates that the decay length of polaritons who are more excitonic is significantly shorter than that of polaritons who are more photonic and that there might be a redistribution of polaritons. This effect is negligible for the case where $F=0$. Such loss and redistribution of populations seem consistent with the picture of the dominant role of dipolar interactions in dynamical scattering processes. To model such dynamics, we write a rate equation for the total propagating polariton population, following their initial excitation:
\begin{equation}\label{Eq:differential}
 \frac{\partial n}{\partial t}=-\gamma_{pol}\cdot n-\gamma_{col}\cdot n^2
\end{equation}
where $\gamma_{pol}$ is the single particle loss rate  and $\gamma_{col}$ is the coefficient of collisions between two polaritons. The second term in Eq.\ref{Eq:differential} represents the loss due to polariton-polariton collisions which, as discussed above, result in scattering outside of the guided modes and thus in loss of WG-polaritons \footnote{we exclude the possibility of a self diffusive motion of the polaritons, as unlike the case of polariton diffusion in slab waveguides \cite{zaitsev_diffusive_2015}, here any scattering-induced in-plane momentum change that is larger than the small lateral  critical angle of the WG mode would result in an in-plane scattering out of the WG and an effective loss rather than diffusive motion}. When divided by the group velocity $v_g^{pol}$, Eq.\ref{Eq:differential} can be rearranged to give the spatial decay along the WG. This differential equation can be solved analytically giving: 
\begin{equation}\label{Eq:solution}
 n(x)=\frac{e^{-x/L}}{n_0^{-1}+\frac{\gamma_{col}}{\gamma_{pol}}(1-e^{-x/L})}
\end{equation}
Here $L=v_g^{pol}/\gamma_{pol}$ is the single polariton decay length and $n_0$ is the initial density. In Figure \ref{fig:fig3}(b) we plot the relative change in the total emission intensity of the polaritonic branch as a function of $\Delta X$ for four different values of $F$ together with the fits to Eq.\ref{Eq:solution}. For $F=0$, the decay fits well to an exponent with $\gamma_{col}\simeq 0$ indicating negligible 2-body collisions. Clearly, the decay length decreases as $F$ increases and the decay becomes non-exponential. In Figure \ref{fig:fig3}(c) we plot the extracted values of $\gamma_{col}$ against $d_0$. It fits well to a quadratic increase with $d_0$, another strong indication of a dipole-dipole induced scattering \cite{bohn_quasi-universal_2009}.

We have established the dominant role dipole-dipole interactions take in the dynamics of a WG-dipolariton system, overwhelming the short-range contact interactions of unpolarized polaritons by orders of magnitude. Surprisingly, the absolute value of $g_d=\alpha d$ is about three orders of magnitude larger than the value theoretically predicted by a mean-field model of interacting dipoles which suggests $\alpha=g_d/d=4\pi e^2/\epsilon=1.4$ $meV\mu m$ \cite{laikhtman_exciton_2009}. This is a significant discrepancy. We note that such a difference is similar to the one reported previously between experiment and theory in the case of the values of $g_0$, where experiments reported $g_0$ values that were about two orders of magnitude above the theoretical predictions\cite{walker_dark_2017,noauthor_direct_nodate}. This fact might hint that the origin of these discrepancies is the same in both cases (dipolar and non-dipolar).  Note that our extracted value of $g_0$ is quite similar with that reported by Walker et al.\cite{walker_dark_2017} in a similar system, which also points toward that conclusion, and also fortifies the validity of our careful density calibration (see details in the supplemental material).


One possibility for such large blue-shifts at the readout point is if they are "carried" from the high density regime at the excitation point without energy relaxation. However, such energy-conserving motion would imply the following scenario: as the polaritons move away from the large density zone to the low density readout area, their initially blue-shifted dispersion should red-shift, but due to energy conservation, the polaritons should "traverse" up along the \textit{less} blue-shifted dispersion curve. In contrast to this picture,in our experiment the polariton moves down along the dispersion curve. The fact that the entire branch seems to shift is evidence that the interactions accompany the polaritons as they fly away from the excitation point. Furthermore, the observed decay of population along the way due to dipolar scattering seems to contradict conserved energy transport. To demonstrate that $\Delta E$ results from the local density at the readout point, in Figure  \ref{fig:fig4}(c,d) we plot $\Delta E(n)$ at the readout point for the two types of experiments\footnote{The fact that a large portion of the more excitonic polaritons are absent at long distances due to efficient scattering, makes it difficult to fit the coupled oscillator model correctly as we increase the excitation distance. For this reason the comparison in Figure \ref{fig:fig4} is made by following the energy change of polaritons with a specific $\beta=\beta_a$ instead.}. In the first experiment we fixed the excitation power at the excitation point (and thus the density there, $n_{exc}$ and presumably $\Delta E_{exc}$), and vary $\Delta X$. In the second we fixed $\Delta X$ but vary the excitation power (and $n_{exc}$ and $\Delta E(n_{exc})$ correspondingly). The two experiments give almost identical results. This shows that $\Delta E$ is a result of local interactions. 
Another option is the existence of a high, steady-state "hidden" density of excitons or charge carriers that accumulate at the out-coupler area, but do not emit. In order to induce the measured blue-shifts, this density should be three orders of magnitude above the measured polariton density, even at measured distances of hundreds of microns away from the excitation point and should have lifetimes larger than 5$\mu$S, which is the time between excitation pulses. Also, they should have the same density dependence on power and distance as the polaritons, to explain the similarity between the two curves in Figure  \ref{fig:fig4}, and should have a suppressed emission. While this is not strictly impossible, a scenario giving such incidental behavior is unlikely.
Finally, large measured $g_d$ values might be expected if there is a true enhancement of the polaritonic interactions over that of bare excitons. One speculation is that this could arise due to the very low effective mass of the polaritons resulting in very large thermal velocities. These large thermal velocities can perhaps increase the rate of scattering events, and allow polaritons to overcome repulsion and have a closer approach between particles, thus increasing the average repulsive energies they experience during a scattering event. Another option is that polaritons form microscopic high-density droplets, so that local densities within each droplet is much higher than the average polariton density.

The large enhancement of the dipolar interaction of dipolaritons compared to that of unpolarized polaritons could facilitate observation of two-polariton interactions in readily achieved conventional optical confinement (in contrast to deep sub-wavelength optical confinement), leading the way towards polariton-based quantum gates\cite{Blockade}. The observed orders of magnitude enhancement of the polariton-polariton interactions induced by their electrical dipole, together with the ability to precisely control this enhancement and effectively turn it on and off via the external bias in well defined and selective locations via patterned electrical gates, can now allow to design complex waveguide structures where the interactions between polaritons can be spatially and temporally controlled. This, together with the long-range nature of dipolar interactions might be used to construct various lattice simulations, and it is such set of properties that were theoretically required for observing a new type of topological states, namely topolaritons in waveguides \cite{Topolaritons}. The ability demonstrated here to truly guide polaritons in fully confined waveguides without any complex sample etching opens up possibilities for polariton-based complex circuitry.


\section*{Acknowledgments}
We would like to acknowledge financial support from the U.S. Department of Energy: Office of Basic Energy Sciences - Division of Materials Sciences and Engineering, from the United State - Israel Binational Science Foundation (BSF grant  2016112), and from the Israeli Science Foundation (grant No. 1319/12). The work at Princeton University was funded by the Gordon and Betty Moore Foundation through the EPiQS initiative Grant GBMF4420, and by the National Science Foundation MRSEC Grant DMR-1420541.
\bibliography{Bibi}
\bibliographystyle{Science}

\includepdf[pages=-,width=1.5\textwidth] {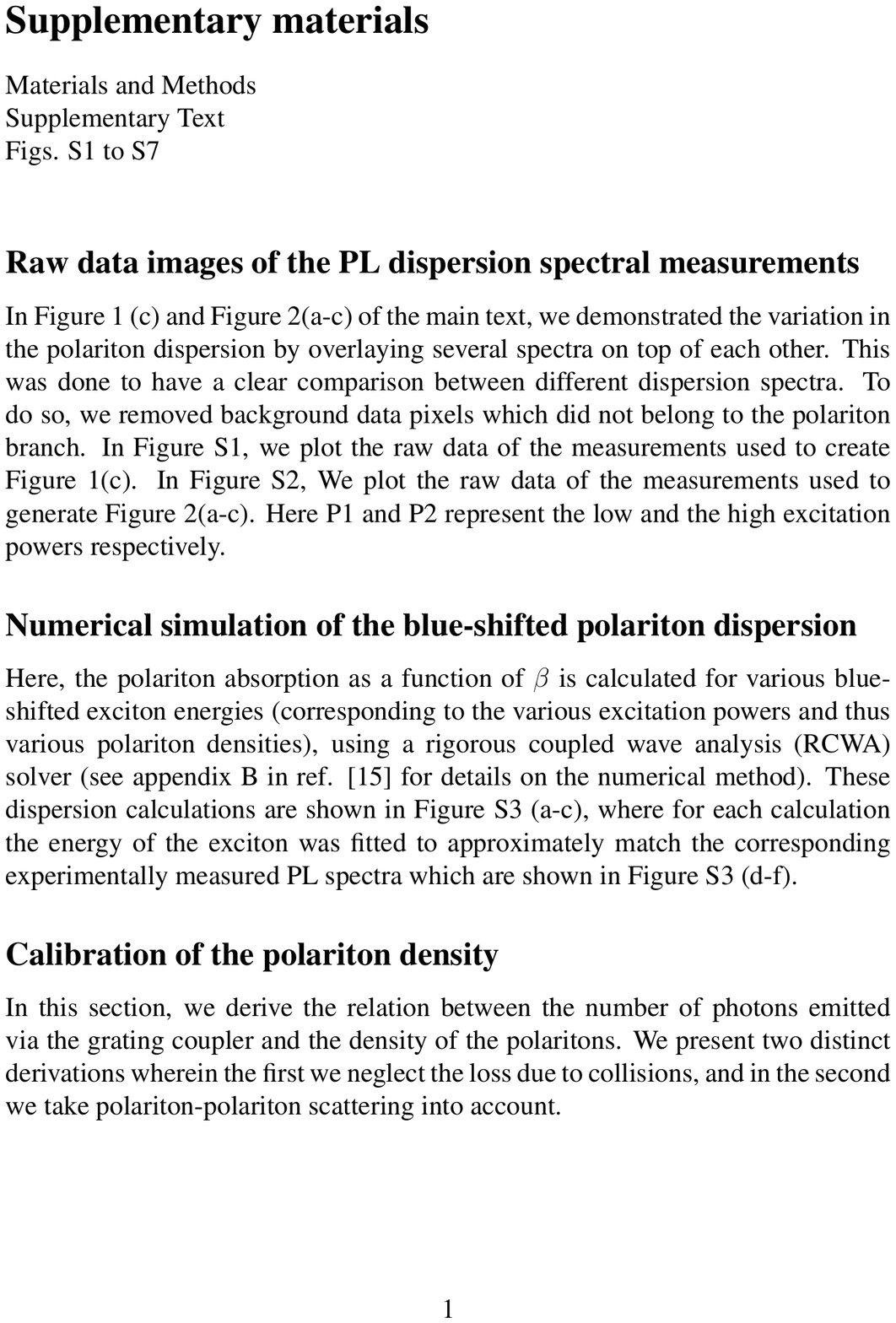}
\end{document}


\section*{Supplementary materials}
Materials and Methods\\
Supplementary Text\\
Figs. S1 to S7\\
\subsection*{Raw data images of the PL dispersion spectral measurements}
\begin{figure}[h!]
  \centering
  \includegraphics[width=0.7\textwidth]{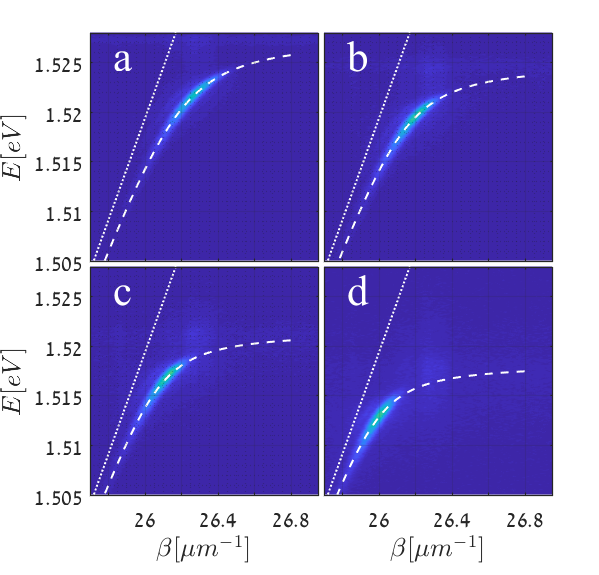}
  \caption{Raw images of the PL dispersion spectral measurements used to create Figure 1(c) of the main text. a) $F=0$. b) $F=11\: kV/cm$. c) $F=17\: kV/cm$. d) $F=22\: kV/cm$.}
    \label{fig:sup:waterfallRaw}
\end{figure}

\begin{figure}[tbp]
  \centering
  \includegraphics[width=0.8\textwidth]{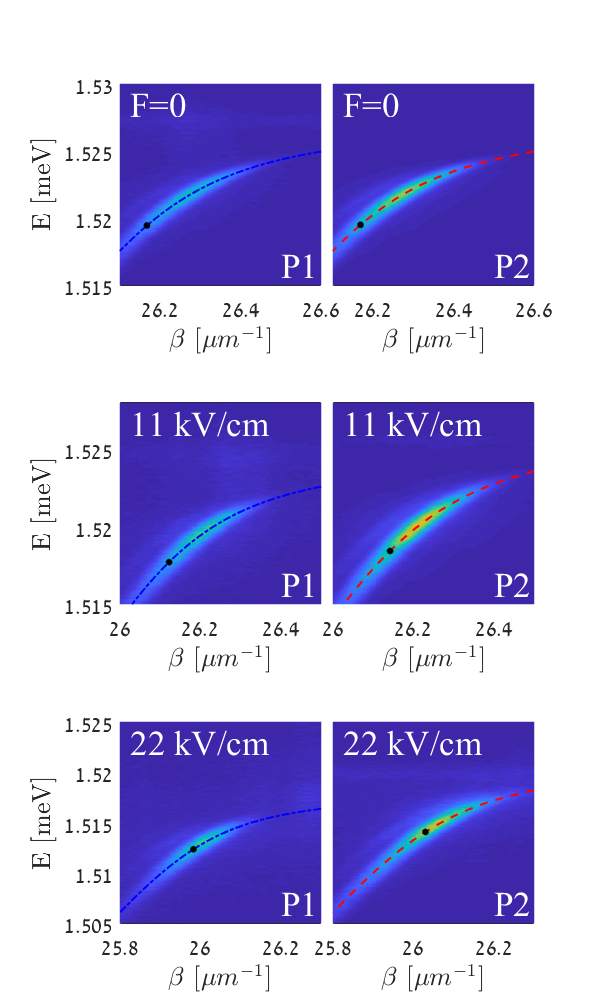}
  \caption{Raw images of the PL dispersion spectral measurements used to create Figure 2(a-c) of the main text. P1 represent low excitation power and P2 represents high excitation power.}
    \label{fig:sup:WeakStrong}
\end{figure}

In Figure 1 (c) and Figure 2(a-c) of the main text, we demonstrated the variation in the polariton dispersion by overlaying several spectra on top of each other. This was done to have a clear comparison between different dispersion spectra. To do so, we removed background data pixels which did not belong to the polariton branch.  In Figure \ref{fig:sup:waterfallRaw},  we plot the raw data of the measurements used to create Figure 1(c). In Figure \ref{fig:sup:WeakStrong}, We plot the raw data of the measurements used to generate Figure 2(a-c). Here P1 and P2 represent the low and the high excitation powers respectively.

\subsection*{Numerical simulation of the blue-shifted polariton dispersion}
\begin{figure}[h!]
  \centering
  \includegraphics[width=0.9\textwidth]{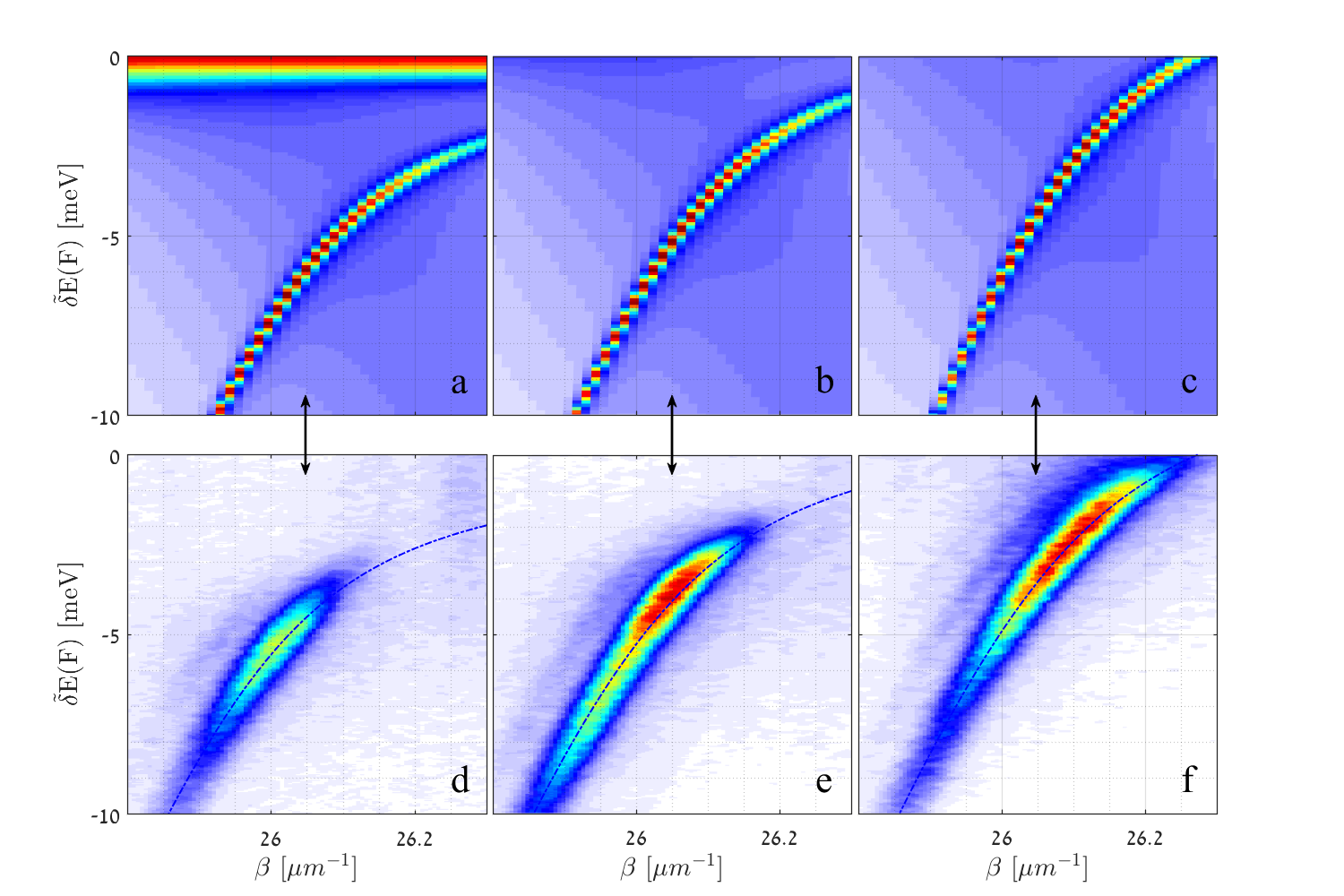}
  \caption{Fitted RCWA simulations of the absorption spectrum as a function of the parallel momentum vector $\beta$, calculated for $F=22 kV/cm$ with fitting exciton blue-shift energies (a) $\Delta E_x=0$ (b) $\Delta E_x=1.5meV$ and (c) $\Delta E_x=3meV$. $\tilde{\delta}$E is the energy with respect to $E_x(\Delta E_X=0)$. (d-f) The corresponding experimental PL spectra.}
    \label{fig:sup:simulated}
\end{figure}

Here, the polariton absorption as a function of $\beta$ is calculated for various blue-shifted exciton energies (corresponding to the various excitation powers and thus various polariton densities), using a rigorous coupled wave analysis (RCWA) solver (see appendix B in ref. [15] for details on the numerical method). These dispersion calculations are shown in Figure \ref{fig:sup:simulated} (a-c), where for each calculation the energy of the exciton was fitted to approximately match the corresponding experimentally measured PL spectra which are shown in Figure \ref{fig:sup:simulated} (d-f).

\subsection*{Calibration of the polariton density}
In this section, we derive the relation between the number of photons emitted via the grating coupler and the density of the polaritons. We present two distinct derivations wherein the first we neglect the loss due to collisions, and in the second we take polariton-polariton scattering into account.

\subsubsection*{Density decay without collisions}
\begin{figure}[h]
  \centering
  \includegraphics[width=0.8\textwidth]{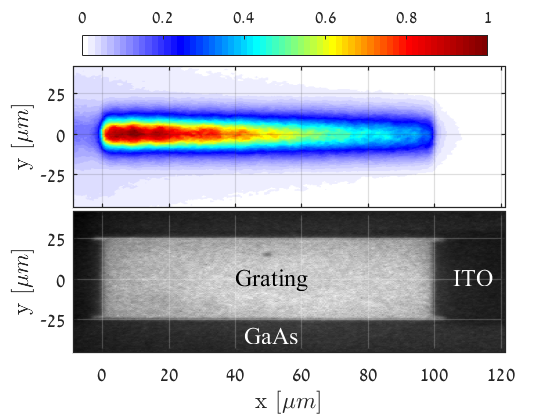}
  \caption{An image of the grating when it is illuminated by white light (bottom) and the normalized PL along the grating, measured for $F=0$ (top). }
    \label{fig:sup:DecayalongGrating1}
\end{figure}

\begin{figure}[h]
  \centering
  \includegraphics[width=0.6\textwidth]{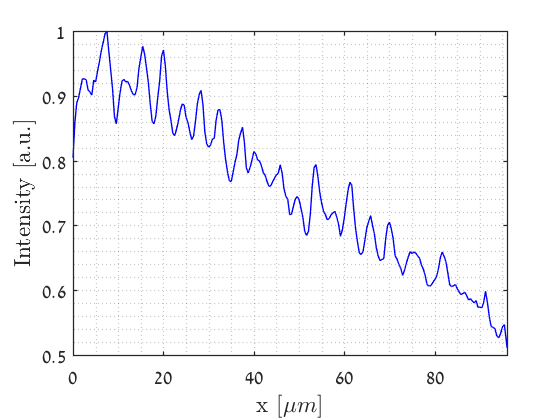}
  \caption{The emission along the grating for $F=0$ }
    \label{fig:sup:DecayalongGrating2}
\end{figure}

\begin{figure}[h]
  \centering
  \includegraphics[width=0.6\textwidth]{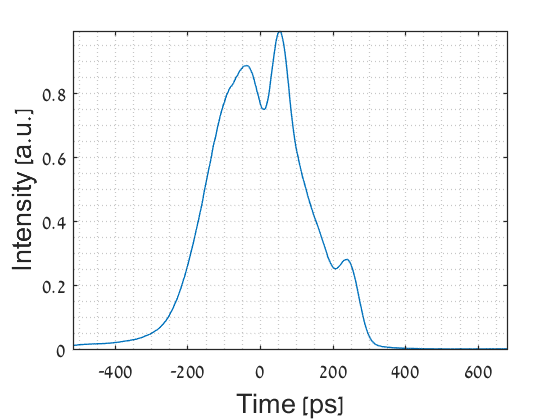}
  \caption{The temporal shape of the laser pulse, measured with a Streak camera. }
    \label{fig:sup:pulse}
\end{figure}

The rate equation for polaritons which propagate along the out-coupling grating is
\begin{equation}\label{Eq:diff_1}
\frac{\partial n}{\partial t}=-\gamma n
\end{equation}
where $\gamma=\gamma_r+\gamma_{nr}$ is the overall decay rate, $\gamma_{nr}$ is the decay along the ITO channel, and $\gamma_r$ is the added loss due to the out-coupling along the grating. In Figure \ref{fig:sup:DecayalongGrating1}  we plot the emitted PL from the area of the grating for $F=0$. The decay along the grating is plotted in Figure \ref{fig:sup:DecayalongGrating2}. Since $\gamma_r$ should not depend on $F$ we can extract: $\gamma_r=\gamma-\gamma_{nr}$.
By dividing both sides of Eq.\ref{Eq:diff_1} by the group velocity of the polaritons $v_g$ we get a differential equation which describes the behavior of the density along the grating.
Solving it gives us the variation of the density along the grating:
\begin{equation}
n(x)=n_0e^{-\tilde{\gamma}x}
\end{equation}
where $\tilde{\gamma}=\gamma/v_g$.
The emitted intensity is given by
\begin{equation}
I(x)=\gamma_rn(x)=v_g\tilde{\gamma}_rn(x).
\end{equation}
The total number of emitted photons which were originated by a single excitation pulse is given by: 
\begin{equation}
N_{ph}=\int_{-\frac{\Delta y}{2}}^\frac{\Delta y}{2}\int_0^{t_p}\int_0^dI(x)dxdtdy
\end{equation}
where $t_p=280\:ps$ is the temporal width of the pulse (see Figure \ref{fig:sup:pulse}), $d=100\:\mu m$ is the length of the grating  and $\Delta y=17\:\mu m$ is the size of the polariton cloud in the perpendicular direction (see Figure \ref{fig:sup:DecayalongGrating1}) thus,
\begin{equation}
N_{ph}=\Delta yv_gt_p\tilde{\gamma}_rn_0\int_0^de^{-\tilde{\gamma}x}=\Delta yv_gt_p\tilde{\gamma}_rn_0(1-e^{-\tilde{\gamma}d})/\tilde{\gamma}
\end{equation}
and
\begin{equation}
n_0=\frac{\tilde{\gamma}N_{ph}}{\Delta yv_gt_p\tilde{\gamma}_r((1-e^{-\tilde{\gamma}d}))}.
\end{equation}
The emission rate $\tilde{\gamma}_r$ was extracted using the data in Figure \ref{fig:sup:DecayalongGrating2} and the data from Figure 4 of the main text, and was found to be $\tilde{\gamma}_r=\tilde{\gamma}-\tilde{\gamma_{nr}}\sim 1/330\:\mu m^{-1}$. 
%
\subsubsection*{Density decay with collisions}
Accounting for collisions, the rate equation will take the form:
\begin{equation}
\frac{\partial n}{\partial t}=-\gamma n-\gamma_{col}n^2
\end{equation}
and the expression for $I(x)$ will be
\begin{equation}
I(x)=\frac{v_g\tilde{\gamma}_re^{-\tilde{\gamma}x}}{n_0^{-1}+\frac{\tilde{\gamma}_{col}}{\tilde{\gamma}}(1-e^{-\tilde{\gamma}x})}.
\end{equation}
Repeating the steps from before we need to solve an integral of the form
\begin{equation}
I_0=\int_0^d\frac{e^{-\tilde{\gamma}x}}{a+b(1-e^{-\tilde{\gamma}x})}dx.
\end{equation}
By defining $\eta=1-e^{-\tilde{\gamma}x}$ we get $d\eta=\tilde{\gamma}e^{-\tilde{\gamma}x}dx$ and the integral becomes
\begin{equation}
I_0=\frac{1}{\tilde{\gamma}}\int_0^{1-e^{-\tilde{\gamma}d}}\frac{d\eta}{a+b\eta}=\frac{1}{\tilde{\gamma}b}log(1+\frac{b}{a}(1-e^{-\tilde{\gamma}d}))
\end{equation}
 in our case $a=n_0^{-1}$, $b=\frac{\tilde{\gamma}_{col}}{\tilde{\gamma}}$ and
\begin{equation}
N_{ph}=\frac{\Delta yv_gt_p\tilde{\gamma}_rlog(1+\frac{n_0\tilde{\gamma}_{col}}{\tilde{\gamma}}(1-e^{-\tilde{\gamma}d}))}{\tilde{\gamma}_{col}}.
\end{equation}
Finally we get:
\begin{equation}\label{Eq:sup:Density}
n_0=\frac{\tilde{\gamma}}{\tilde{\gamma}_{col}(1-e^{-\tilde{\gamma}d})}(e^\frac{N_{ph}\tilde{\gamma}_{col}}{\Delta yv_gt_p\tilde{\gamma}_r}-1).
\end{equation}
\subsubsection*{Density calibration procedure}
The expression in Eq.\ref{Eq:sup:Density} is the one we used to derive the density of the polaritons. Where $t_p$, $\Delta y$, $d$ and $\tilde{\gamma}_r$ were defined in the previous section. $\tilde{\gamma}_{col}$ is the same value as the one calculated by fitting to Eq.4 of the main text. The group velocity, $v_g$, which is the mean velocity of the polaritons, was extracted from the fits to the coupled oscillator model. From the fits, we get the population at each excitonic fraction $\chi_x(\beta)$ (or photonic fraction $\chi_{ph}(\beta)$), so we can find the mean polaritonic fractions and the mean velocity by taking the derivatives of the dispersion.
$N_{ph}$ is acquired from the counts of the ICCD. We used the following routine:
First, we quantified the loss of the optical setup, we have let a laser beam to traverse along the optical setup, and we measured the power before it entered the setup and  before it entered the CCD (a fraction of 0.4 got through).
Then, we measured how many counts we get from the ICCD for a specific amount of radiation energy both at the 0th and the 1st order of the spectrometer. This gave us the conversion between counts to photons which was then used in the experiment.
In addition, we also used an FDTD simulation to see the ratio between the number of photons who got emitted toward the objective and the number of photons who got emitted toward the substrate. We found that the amounts are roughly equal, so we multiplied the number of counts by a factor of 2. 
Since the angular emission from the sample (in the y-direction) is $\sim 30^\circ$ and the slit lets only $0.73^\circ$ to enter we also multiplied the number of counts by a factor of 41.
%
%
%
\subsection*{Extraction of the separate variations of the exciton energy and of $\Omega_{xp}$ on $n$}
\begin{figure}[h!]
  \centering
  \includegraphics[width=1\textwidth]{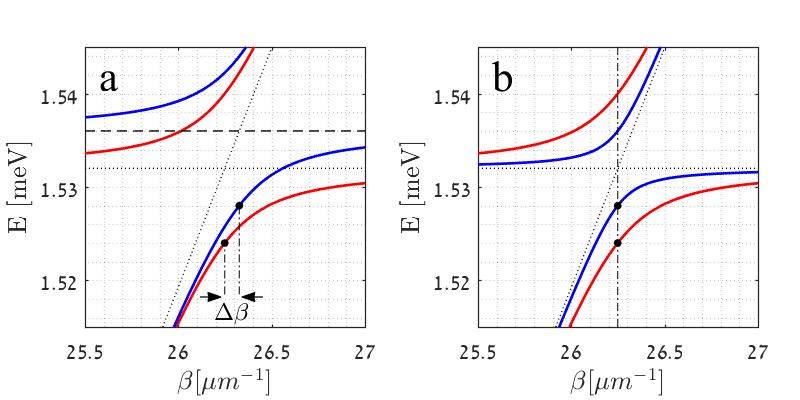}
  \caption{Illustration of the two mechanisms which can blue-shift the energy of the polariton. a) Blue-shift in the energy of the exciton: The black dotted lines are the bare exciton and bare WG-mode and the black dashed line is the energy of the exciton after a blue-shift. The continuous red (blue) lines are the LP and UP modes before (after) the blue shift in the energy of the bare exciton. The black dots mark the energy and momentum of the $\chi_x=1/2$ polariton. b) Reduction in the oscillator strength due to phase space filling or screening of the binding energy: The dotted lines, the continuous lines and the black dots have the same role as in (a). From this figure, it can be understood that only real blue-shift in the energy of the exciton will cause a change in the parallel momentum of the $\chi_x=1/2$ polariton.}
    \label{fig:sup:Blue-shiftOrgiin}
\end{figure}

It is essential to unambiguously determine whether the blue-shift we measure results from an interaction-induced energy shift of the bare exciton or a reduction in the oscillator strength of the exciton. It turns out that due to the particular dispersion of WG-polaritons, these two effects are separable. By following the parallel momentum, $\beta$, of the equal exciton and photon fraction $\chi_x=1/2$ of the lower polariton branch, we can distinguish between these two mechanisms: Since the parallel momentum of the $\chi_x=1/2$  is equal by the definition to the parallel momentum of the  crossing point of the dispersions of the bare exciton and the bare photon, we do not expect to see any change in $\beta(\chi_x=1/2)$ if the blue-shift is solely due to a reduction in the oscillator strength. However, if the blue-shift arises from an actual energy shift of the excitonic component, the crossing point should also shift to larger $\beta$ values. This is depicted schematically in Figure \ref{fig:sup:Blue-shiftOrgiin}. Moreover, by following the variations in $\beta(\chi_x=1/2)$ and plugging it to the dispersion of the bare photon we can derive the magnitude by which the energy of the exciton was shifted and the variation of $\Omega_{xp}$, as was done to extract those values plotted in Figure  2 of the main text.
\subsection*{The rate equation of propagating polaritons in a waveguide}
In this section, we derive the expression describing the rate equation for the decay along the channel (Eq.3 of the main text). We begin by writing a general expression for $n_\beta$, a density of polaritons with a given propagation constant $\beta$. The change in the population of this density is determined by the single particle decay rate $\gamma_\beta$ and by the probability for these polaritons to scatter with polaritons from the overall density. Thus we can write: 
\begin{equation}\label{Eq:rate1}
\frac{\partial n_\beta}{\partial t}=-\gamma_\beta n_\beta-\gamma_{col}^\beta n_\beta n_{tot}.
 \end{equation}
Now $n_\beta$ can be represented as  $n_\beta=P_\beta n_{tot}$, where $n_{tot}$ is the total density and $P_\beta$ is the distribution function of the polaritons. Plugging it in into Eq.\ref{Eq:rate1} and summing over $\beta$ we get
\begin{equation}\label{Eq:rate2}
\sum_\beta\frac{\partial n_\beta}{\partial t}=-(\sum_\beta\gamma_\beta P_\beta) n_{tot}-(\sum_\beta\gamma_{col}^\beta P_\beta) n_{tot} ^2
 \end{equation}
 Now we define $\gamma_{pol}=\sum_\beta\gamma_\beta P_\beta$ and $\gamma_{col}=\sum_\beta\gamma_{col}^\beta P_\beta$ to get Eq.3. 
 \subsection*{Derivation of $d(n)$}
Here we give the derivation of the expression for $d(n)$ used during the fitting of the data in the paper.  Considering the differential of the expression for the blue shift $\Delta E=\alpha d(n)n$ we get
 \begin{equation}\label{Eq:d(n)1}
\delta\Delta E=\alpha d \delta n+\alpha n\frac{\delta d}{\delta n}\delta n.
 \end{equation}
 We assume that $d$ decrease linearly with the increase of the bluse shift $\delta d=\xi \delta\Delta E$ and plug it into Eq.\ref{Eq:d(n)1} to get
 \begin{equation}\label{Eq:d(n)2}
\frac{\delta d}{\xi}=\alpha d \delta n+n\alpha \frac{\delta d}{\delta n}\delta n.
 \end{equation}
Eq.\ref{Eq:d(n)3} can be rearranged to give
\begin{equation}\label{Eq:d(n)3}
\frac{\delta d}{\delta n}=\frac{\xi\alpha d}{1-\xi\alpha n}
\end{equation}
Which by integration gives the solution
\begin{equation}\label{Eq:d(n)4}
d(n)=\frac{d_0}{1-\alpha\xi n}.
\end{equation}
\subsection*{The expression for the energy variation at constant $\beta$}
In Figure 3(c,d) of the main text, we show the variation of the blue-shift at a specific parallel momentum. It can be noticed that the values of the blue-shift there are smaller than those we got for polaritons with a specific excitonic fraction. In the following, we offer a derivation which gives the expression for the variation in energy for constant $\beta$. We note that this is usually the case one encounters when regarding the density-dependent blue-shift in microcavities, where $\beta=0$. 
The energy of the lower polariton branch is given by
\begin{equation}
E(\beta)_{pol}=\frac{1}{2}\left[E_x(\beta)+E_{ph}(\beta)-\sqrt{(E_{ph}(\beta)-E_x(\beta))^2+4\hbar^2\Omega^2}\right].
\end{equation}
Now lets assume we have some density of dipoles in the system $\bar{n}$ which introduces a blue-shift to the excitons energy $E_x\rightarrow E_x+\bar{g}\bar{n}$.  In general $\bar{n}=n_{pol}\chi_x(\beta)$ and $\bar{g}=g_0$ or $\bar{g}=g_d(n)$.\\
The blue-shift of the polariton branch at constant $\beta$ will be given by
\begin{equation}
 \begin{array}{cc}
\Delta E(\beta)=\frac{1}{2}\left[E_x(\beta)+\bar{g}\bar{n}+E_{ph}(\beta)-\sqrt{(E_{ph}(\beta)-E_x(\beta)-\bar{g}\bar{n})^2+4\hbar^2\Omega^2}\right] &  \\
 & \\
- & \\
& \\
\frac{1}{2}\left[E_x(\beta)+E_{ph}(\beta)-\sqrt{(E_{ph}(\beta)-E_x(\beta))^2+4\hbar^2\Omega^2}\right] &  \\
& \\
= & \\
 & \\
\frac{1}{2}\left[\bar{g}\bar{n}+\sqrt{(E_{ph}(\beta)-E_x(\beta))^2+4\hbar^2\Omega^2}-\sqrt{(E_{ph}(\beta)-E_x(\beta)-\bar{g}\bar{n})^2+4\hbar^2\Omega^2}\right].
 \end{array} 
\end{equation}
We define $\delta(\beta)=E_{ph}(\beta)-E_x(\beta)$ and get
\begin{equation}
\Delta E(\beta)=\frac{1}{2}\left[\bar{g}\bar{n}+\sqrt{(\delta(\beta))^2+4\hbar^2\Omega^2}-\sqrt{(\delta(\beta)-\bar{g}\bar{n})^2+4\hbar^2\Omega^2}\right]
\end{equation}
which can be further rearranged as
\begin{equation}
\Delta E(\beta)=\frac{1}{2}\left[\bar{g}\bar{n}+\sqrt{(\delta(\beta))^2+4\hbar^2\Omega^2}-\sqrt{(\delta(\beta))^2+4\hbar^2\Omega^2 -2\delta(\beta)\bar{g}\bar{n}+(\bar{g}\bar{n})^2}\right].
\end{equation}
Assuming $\bar{g}\bar{n}<(\delta(\beta))^2+4\Omega^2)$ we can approximate:
\begin{equation*}
\Delta E(\beta)=\frac{1}{2}\left[\bar{g}\bar{n}+\sqrt{(\delta(\beta))^2+4\hbar^2\Omega^2}-\sqrt{(\delta(\beta))^2+4\hbar^2\Omega^2}\sqrt{1 -\frac{2\delta(\beta)\bar{g}\bar{n}-(\bar{g}\bar{n})^2}{(\delta(\beta))^2+4\hbar^2\Omega^2}}\right]
\end{equation*}
\begin{equation*}
\Delta E(\beta)\approx\frac{1}{2}\left[\bar{g}\bar{n}+\sqrt{(\delta(\beta))^2+4\hbar^2\Omega^2}-\sqrt{(\delta(\beta))^2+4\hbar^2\Omega^2}(1 -\frac{2\delta(\beta)\bar{g}\bar{n}-(\bar{g}\bar{n})^2}{2((\delta(\beta))^2+4\hbar^2\Omega^2))}\right]
\end{equation*}
\begin{equation}\label{Eq:BS}
\Delta E(\beta)=\frac{1}{2}\left[\bar{g}\bar{n}+ \frac{\bar{g}\bar{n}\delta(\beta)}{\sqrt{\delta(\beta))^2+4\hbar^2\Omega^2}}-\frac{(\bar{g}\bar{n})^2}{2\sqrt{\delta(\beta))^2+4\hbar^2\Omega^2}}\right].
\end{equation}
Finally we can use that fact that:
\begin{equation}
\chi_x(\beta)=\frac{1}{2}\left[1+\frac{\delta(\beta)}{\sqrt{\delta(\beta))^2+4\hbar^2\Omega^2}}\right]
\end{equation}
and get
\begin{equation}\label{Eq:last}
\Delta E(\beta)=\bar{g}\bar{n}\left[\chi_x(\beta)-\frac{\bar{g}\bar{n}}{4\sqrt{(\delta(\beta))^2+4\hbar^2\Omega^2}}\right]. 
\end{equation}
It can be seen that the factor $\bar{g}\bar{n}$ is multiplied by the excitonic fraction. At very low densities the second term of Eq.\ref{Eq:last} can be neglected and the blue-shift of the polariton will be the blue-shift of the exciton  multiplied by the excitonic fraction.
When neglecting the second term in the above equation, the ratio between the variation in energy of polaritons with constant $\beta$ and the blue shift of the $\chi_x=1/2$ polariton is given by:
\begin{equation}
\frac{\Delta E(\beta)}{\Delta E(\chi_x=1/2)}\sim\frac{\bar{g}\bar{n}\chi_x(\beta)}{\bar{g}\bar{n}}=\chi_x(\beta)
\end{equation}
which explains the difference between the two magnitudes.


\clearpage
